\documentclass[aps,prl,preprint,superscriptaddress,showpacs]{revtex4}
\usepackage{graphicx}
\usepackage{dcolumn}

\begin{document}

\def\GSI{GSI Helmholtzzentrum f{\"u}r Schwerionenforschung GmbH, D-64291 Darmstadt, Germany}
\def\IPNO{Institut de Physique Nucl{\'e}aire, IN2P3-CNRS et Universit{\'e}, F-91406
Orsay, France}
\def\JAGU{M. Smoluchowski Institute of Physics, Jagiellonian University, Pl-30059 Krak\'ow, Poland}
\def\MILANO{Istituto di Scienze Fisiche, Universit\`a degli Studi and INFN, I-20133 Milano, Italy}
\def\MOSCOW{Institute for Nuclear Research, 117312 Moscow, Russia}
\def\LPC{LPC, IN2P3-CNRS, ISMRA et Universit{\'e}, F-14050 Caen, France}
\def\SACLAY{DAPNIA/SPhN, CEA/Saclay, F-91191 Gif-sur-Yvette, France}
\def\GANIL{GANIL, CEA et IN2P3-CNRS, F-14076 Caen, France}
\def\IFJ{IFJ-PAN, Pl-31342 Krak\'ow, Poland}
\def\CATANIA{Dipartimento di Fisica e Astronomia dell'Universit\`a and INFN-LNS, 
I-95123 Catania, Italy}
\def\MSU{Department of Physics and Astronomy and NSCL, MSU, East Lansing, MI 48824, USA}
\def\WARSAW{A.~So{\l}tan Institute for Nuclear Studies, Pl-00681 Warsaw, Poland}

\title{Isotopic Dependence of the Nuclear Caloric Curve
}

\affiliation{\GSI}
\affiliation{\IPNO}
\affiliation{\JAGU}
\affiliation{\MILANO}
\affiliation{\MOSCOW}
\affiliation{\SACLAY}
\affiliation{\GANIL}
\affiliation{\IFJ}
\affiliation{\CATANIA}
\affiliation{\MSU}
\affiliation{\WARSAW}

\author{C.~Sfienti}             \affiliation{\GSI}
\author{P.~Adrich}              \affiliation{\GSI}  
\author{T.~Aumann}              \affiliation{\GSI} 
\author{C.O.~Bacri}             \affiliation{\IPNO} 
\author{T.~Barczyk}             \affiliation{\JAGU} 
\author{R.~Bassini}             \affiliation{\MILANO}  
\author{S.~Bianchin}            \affiliation{\GSI} 
\author{C.~Boiano}              \affiliation{\MILANO}   
\author{A.S.~Botvina}           \affiliation{\GSI}\affiliation{\MOSCOW}
\author{A.~Boudard}             \affiliation{\SACLAY}  
\author{J.~Brzychczyk}          \affiliation{\JAGU}   
\author{A.~Chbihi}              \affiliation{\GANIL}
\author{J.~Cibor}               \affiliation{\IFJ}
\author{B.~Czech}               \affiliation{\IFJ} 
\author{M.~De~Napoli}           \affiliation{\CATANIA}
\author{J.-\'{E}.~Ducret}       \affiliation{\SACLAY}
\author{H.~Emling}              \affiliation{\GSI}
\author{J.D.~Frankland}           \affiliation{\GANIL}
\author{M.~Hellstr\"{o}m}       \affiliation{\GSI}
\author{D.~Henzlova}            \affiliation{\GSI}
\author{G.~Imm\`{e}}            \affiliation{\CATANIA} 
\author{I.~Iori}\thanks{deceased} \affiliation{\MILANO}  
\author{H.~Johansson}           \affiliation{\GSI} 
\author{K.~Kezzar}              \affiliation{\GSI}
\author{A.~Lafriakh}            \affiliation{\SACLAY}
\author{A.~Le~F\`evre}          \affiliation{\GSI}
\author{E.~Le~Gentil}           \affiliation{\SACLAY}
\author{Y.~Leifels}             \affiliation{\GSI}
\author{J.~L\"{u}hning}         \affiliation{\GSI}
\author{J.~{\L}ukasik}          \affiliation{\GSI}\affiliation{\IFJ} 
\author{W.G.~Lynch}             \affiliation{\MSU}
\author{U.~Lynen}               \affiliation{\GSI} 
\author{Z.~Majka}               \affiliation{\JAGU}  
\author{M.~Mocko}               \affiliation{\MSU}
\author{W.F.J.~M\"{u}ller}      \affiliation{\GSI}
\author{A.~Mykulyak}            \affiliation{\WARSAW}          
\author{H.~Orth}                \affiliation{\GSI}
\author{A.N.~Otte}              \affiliation{\GSI}
\author{R.~Palit}               \affiliation{\GSI}
\author{P.~Paw{\l}owski}        \affiliation{\IFJ}
\author{A.~Pullia}              \affiliation{\MILANO}
\author{G.~Raciti}              \affiliation{\CATANIA}
\author{E.~Rapisarda}           \affiliation{\CATANIA} 
\author{H.~Sann}\thanks{deceased} \affiliation{\GSI}
\author{C.~Schwarz}             \affiliation{\GSI}
\author{H.~Simon}               \affiliation{\GSI}
\author{K.~S\"{u}mmerer}        \affiliation{\GSI}
\author{W.~Trautmann}           \affiliation{\GSI}
\author{M.B.~Tsang}             \affiliation{\MSU}
\author{G.~Verde}               \affiliation{\MSU}
\author{C.~Volant}              \affiliation{\SACLAY} 
\author{M.~Wallace}             \affiliation{\MSU}
\author{H.~Weick}               \affiliation{\GSI}
\author{J.~Wiechula}            \affiliation{\GSI}
\author{A.~Wieloch}             \affiliation{\JAGU} 
\author{B.~Zwiegli\'{n}ski}     \affiliation{\WARSAW}
\collaboration{The ALADIN2000 Collaboration}

\date{\today}

\begin{abstract}

The $A/Z$ dependence of projectile fragmentation at relativistic 
energies has been studied with the ALADIN forward spectrometer at SIS. 
A stable beam of $^{124}$Sn and radioactive beams of $^{124}$La 
and $^{107}$Sn at 600 MeV per nucleon have been used in order to 
explore a wide range of isotopic compositions.
Chemical freeze-out temperatures are found to be nearly invariant with respect 
to the $A/Z$ of the produced spectator sources, consistent with
predictions for expanded systems. Small Coulomb effects ($\Delta T \approx 0.6$~MeV)
appear for residue production near the onset of multifragmentation.

\end{abstract}


\maketitle

The isotopic dependence of the nuclear caloric curve,
the temperature-energy relation of excited nuclear systems ~\cite{poch95,kelic06},
is of interest for several reasons. It is, at first, 
of practical importance for isotopic reaction
studies, presently conducted in many laboratories and conceived under 
the assumption that the basic reaction processes remain unchanged if 
only the isotopic composition of the collision partners is varied. One expects
that specific effects related to the isotopic dependence of the nuclear forces can be
isolated in this way~\cite{colonna06,baoan08}. For example,
in the statistical interpretation of isoscaling, analytic relations 
between the measured parameters and the symmetry term in the equation of state can be
derived if
the freeze-out temperatures can be assumed to be identical for the reactions one 
compares~\cite{colonna06,botv02}. 
A significant isotopic dependence of the caloric curve would here present a complication.

From a theoretical point of view, the isotopic behaviour of the caloric curve is useful 
for investigating its connection with limiting temperatures, i.e. the maximum temperatures  
nuclei can sustain before they become unbound~\cite{bonche85,besp89}. 
These limiting temperatures have been found to be correlated with the critical 
temperature of nuclear matter, in fact nearly linearly in mean-field calculations with Skyrme 
forces~\cite{song91}. Experimental information on limiting temperatures will thus permit tests 
of microscopic calculations of the nuclear equation of state at finite temperature which 
cannot be easily obtained by other means~\cite{baldo04,wang05}.

In the calculations considering excited compound nuclei in equilibrium with their surrounding
vapor, their stability was found to be strongly 
dependent on the Coulomb pressure generated by the protons they contain~\cite{bonche85,besp89}. 
The limiting temperatures decrease along the valley of $\beta$ stability because the effect 
of the increasing atomic number $Z$ is stronger than that of the decreasing charge-to-mass 
ratio $Z/A$ of heavy nuclei. 
A systematic mass dependence of measured breakup temperatures in multifragmentation reactions 
has, therefore, led to the suggestion that they may be identified with the predicted stability 
limits~\cite{nato95,nato02}. In this case, 
since Coulomb effects should be even more pronounced along chains of isotopes 
or isobars, one would expect a significant isotopic dependence of the caloric curve~\cite{kelic06}.

On the other hand, statistical models for multifragmentation, based on calculating the accessible
phase space in expanded volumes~\cite{gross86,bond95}, predict only small temperature
differences for neutron-rich and neutron-poor systems~\cite{ogul02}.
To some extent, the isotopic 
behaviour of the caloric curve thus turns into a test of the reaction mechanism,
indicating whether the observed disintegrations are
primarily caused by a Coulomb instability limiting the existence of compound nuclei or by 
the opening of the partition space. 

Experiment S254, conducted at the SIS heavy-ion synchrotron 
at GSI Darmstadt, was devoted to the study of isotopic effects in projectile 
fragmentation at relativistic energies. Besides stable $^{124}$Sn beams,
neutron-poor secondary Sn and La beams were used in order 
to extend the range of isotopic compositions beyond that available with stable 
beams alone. The radioactive beams were produced  at the fragment 
separator FRS \cite{frs92} by fragmenting primary $^{142}$Nd 
projectiles with energies near 900 MeV/nucleon in a thick beryllium  target. 
The FRS was set to select $^{124}$La and, subsequently,
$^{107}$Sn projectiles which were  
directed onto $^{\rm nat}$Sn targets of 500 mg/cm$^2$ areal density at the ALADIN setup.
All three beams had a laboratory energy of 600 MeV/nucleon. 
At this energy, the acceptance of the ALADIN forward spectrometer 
is about 90\% for projectile fragments with $Z=3$, increases with $Z$, and exceeds
95\% for $Z \ge 6$~\cite{schuett96}.

\begin{figure}[htb]     

   \centerline{\includegraphics[height=7.0cm]{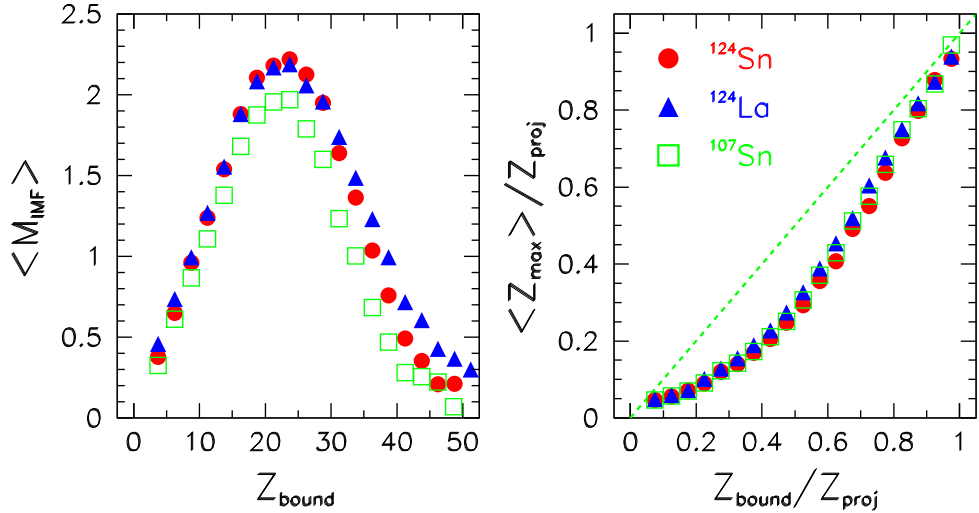}}

\caption{(color online) Acceptance corrected mean multiplicity $<$$M_{\rm IMF}$$>$ of projectile 
fragments for $^{124}$Sn (circles), $^{124}$La (triangles), and $^{107}$Sn
(open squares) beams of 600 A MeV on $^{\rm nat}$Sn targets as a function of $Z_{\rm bound}$ 
(left panel) and correlations of $<$$Z_{\rm max}$$>$ 
with $Z_{\rm bound}$ (both normalized with respect to the atomic number $Z_{\rm proj}$
of the projectile, right panel). 
}
\label{global} 
\end{figure}

In order to reach the necessary beam intensity of about 10$^3$ particles/s with
the smallest possible mass-to-charge ratio $A/Z$, it was found necessary to
accept a distribution of neighbouring nuclides together with the requested
$^{124}$La or $^{107}$Sn isotopes. 
The mean compositions of the nominal $^{124}$La ($^{107}$Sn) beams 
were $<$$Z$$>$ = 56.8 (49.7) and $<$$A/Z$$>$ = 2.19 (2.16), respectively~\cite{luk08}. 
Model studies confirm that these $<$$A/Z$$>$ values are also representative for the 
spectator systems emerging after the initial stages of the reaction~\cite{botv02,lef05}.

The obtained mass resolution 
is about 7\% (FWHM) for projectile fragments with $Z \le 3$ and decreases to 3\% for $Z\geq 6$.
Masses are thus individually resolved for fragments with atomic number $Z \leq 10$.
The elements are individually resolved over the full range of atomic numbers up 
to $Z_{\rm proj}$ with the resolution $\Delta Z \leq 0.6$ (FWHM) obtained with the
TP-MUSIC IV detector~\cite{schuett96}.

Global fragmentation observables were found to depend only weakly on the isotopic composition. 
This is shown in Fig.~\ref{global} for the mean
multiplicity of intermediate-mass fragments ($3 \le Z \le 20$)
and for $Z_{\rm max}$, both as a function of $Z_{\rm bound}$. Here
$Z_{\rm max}$ denotes the largest atomic number $Z$ within a partition while 
the sorting variable $Z_{\rm bound} = \Sigma Z_i$ with $Z_i \ge 2$ 
represents the $Z$ of the spectator system, apart from emitted hydrogens.

The multiplicities exhibit the universal
rise and fall of fragment production \cite{schuett96},
and only a slightly steeper slope in the rise section ($Z_{\rm bound} > 25$)
distinguishes the neutron-rich $^{124}$Sn from the two other cases. 
The difference can be related 
to the evaporation properties of excited heavy nuclei \cite{sfienti_prag}. 
Neutron emission as the prevailing deexcitation mode of neutron-rich residue nuclei
does not affect $Z_{\rm bound}$. The emission of hydrogen isotopes reduces $Z_{\rm bound}$ 
since they are not counted therein.
The same effect produces small differences in the 
correlation of $<$$Z_{\rm max}$$>$  with $Z_{\rm bound}$ (Fig.~\ref{global}, right panel). 
There, the transition from predominantly residue production 
to multifragmentation appears as a reduction of $<$$Z_{\rm max}$$>$ with
respect to $Z_{\rm bound}$ 
which occurs between $Z_{\rm bound}/Z_{\rm proj} =$~0.6 and 0.8.

\begin{figure}[htb]     

   \centerline{\includegraphics[height=7.0cm]{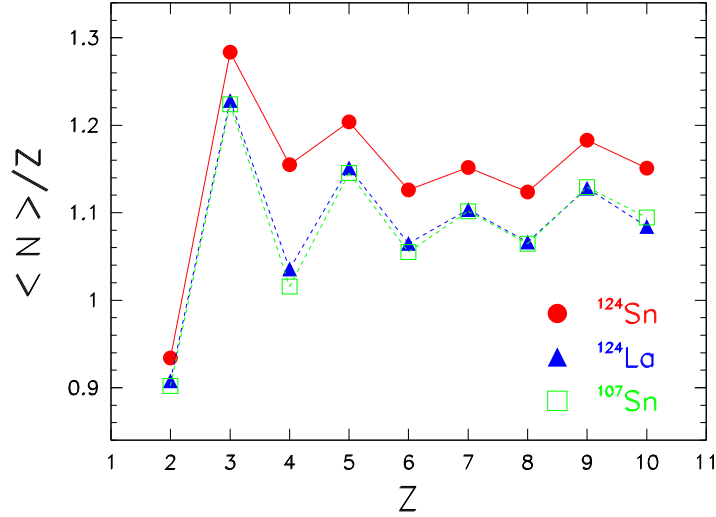}}

\caption{(color online) Mean neutron-to-proton ratios $<$$N$$>$/$Z$ of light fragments 
up to $Z = 10$ 
for $0.2 < Z_{\rm bound}/Z_{\rm proj} \le 0.4$ as a function of the fragment $Z$. 
}
\label{noverz} 
\end{figure}

The mean neutron-to-proton ratios $<$$N$$>$/$Z$ of light fragments
exhibit nuclear structure effects characteristic for the isotopes produced 
as well as a significant memory of the isotopic composition of the emitting 
system (Fig.~\ref{noverz}). 
The mean neutron numbers are larger for the fragments of $^{124}$Sn by, on average, 
$\Delta N = 0.4$.
The values for $Z=4$ have been corrected for the missing yields of unstable $^8$Be fragments
by smoothly interpolating over the measured yields of $^{7,9-11}$Be. 
This has a negligible effect for $^{124}$La and $^{107}$Sn with $<$$N$$>$/$Z \approx 1$
but lowers the value for $^{124}$Sn from 1.23 to 1.16 which makes the systematic odd-even 
variations as a function of the fragment $Z$ more clearly visible.
Apparently, the strongly bound even-even $N=Z$ nuclei attract a large fraction 
of the product yields~\cite{ricci04}. 

\begin{figure}[htb]     

   \centerline{\includegraphics[height=8.0cm]{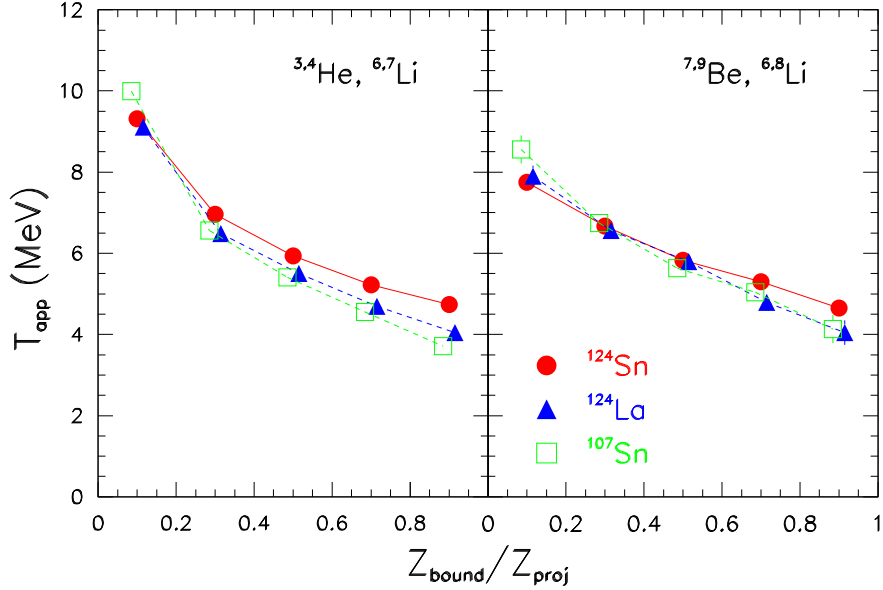}}

\caption{(color online) Apparent temperatures $T_{\rm HeLi}$ (left panel) and $T_{\rm BeLi}$ 
(right panel) as a function of $Z_{\rm bound}/Z_{\rm proj}$ for the three reaction systems.
For clarity, two of the three data sets are slightly
shifted horizontally, only statistical errors are displayed.
}
\label{temp} 
\end{figure}

Two temperature observables, deduced from the resolved isotope yields, 
are shown in Fig.~\ref{temp} as a function of the normalized $Z_{\rm bound}$. 
Besides the frequently used
\begin{equation}
T_{\rm HeLi} = 13.3 MeV/\ln(2.2\frac{Y_{^{6}{\rm Li}}/Y_{^{7}{\rm Li}}}
{Y_{^{3}{\rm He}}/Y_{^{4}{\rm He}}})
\label{eq:heli}
\end{equation}
(left panel, Ref.~\cite{poch95}),
also 
\begin{equation}
T_{\rm BeLi} = 11.3 MeV/\ln(1.8\frac{Y_{^{9}{\rm Be}}/Y_{^{8}{\rm Li}}}
{Y_{^{7}{\rm Be}}/Y_{^{6}{\rm Li}}})
\label{eq:beli}
\end{equation}
deduced from Li and Be fragment yields is displayed (right panel, Ref.~\cite{traut07}). 
The apparent temperatures, as given by the formulae, are shown,
i.e. without corrections for secondary decays feeding the ground states of these nuclei.
Including such corrections will raise the temperature values by 
10 to 20\%~\cite{poch95,traut07}. 
The dependence of the secondary-decay corrections 
on $A/Z$ has been quantitatively studied with two models, the SMM~\cite{bond95} 
and the Quantum Statistical Model 
of Hahn and St{\"o}cker \cite{hahn88} but significant effects ($> 300$~keV) were not found.

Both temperature observables show the same smooth rise with increasing centrality that
is familiar from earlier studies of $^{197}$Au fragmentations~\cite{traut07,xi97}. 
The dependence on the isotopic composition is rather weak. 
The mean temperature differences 
between the neutron-rich and neutron-poor systems amount to 
$\Delta T_{\rm HeLi} = 0.5 \pm 0.1$~MeV and $\Delta T_{\rm BeLi} = 0.1 \pm 0.1$~MeV 
in the bin of maximum fragment production, $Z_{\rm bound}/Z_{\rm proj} \approx 0.5$, 
and become negligible at smaller $Z_{\rm bound}$. This translates into a similar invariance 
for the nuclear caloric curve as $Z_{\rm bound}$ may serve as a measure of the deposited 
energy $E_x$. 
The latter is expected on the basis of the participant-spectator geometry but also 
confirmed by the observation that the degree of fragmentation, known to depend on 
$E_x$~\cite{tamain06}, is strongly related to $Z_{\rm bound}$~(Fig.~\ref{global}). 
At larger $Z_{\rm bound}$,  
in the regime of predominantly residue production ($Z_{\rm bound}/Z_{\rm proj} \approx 0.7$
and above), the temperatures of $^{124}$Sn decays exceed those for the neutron-poor systems
by about 0.6 MeV. 

Within these limits, and particularly in the regime of multifragmentation, 
the deduced temperatures are consistent with the overall observation 
that the reaction processes are not strongly affected by a variation of the system $A/Z$.
Only the fragment mass distributions react sensitively to this parameter (Fig.~\ref{noverz}). 
Comparing to the theoretical predictions, we find that the global behavior of the breakup 
temperatures is in good agreement with the SMM calculations for $^{124}$Sn and $^{124}$La 
nuclei of Ogul and Botvina~\cite{ogul02}. The differences obtained for these cases 
are negligible in the 
multifragmentation regime and reach a maximum $\Delta T \approx 0.4$~MeV in the transition 
region where the equilibrium temperature for the more proton-rich $^{124}$La system
is slightly lower.

Rather small isotopic variations of the caloric curve have also been predicted with 
Thomas-Fermi-type calculations in which expansion~\cite{kolomietz01,hoel07,samaddar07} and 
shape degrees of freedom~\cite{de06} have been considered. The obtained temperatures of 
5 to 8~MeV are within the present range
but it is not obvious that the experimental temperatures of fragmented systems at chemical 
freeze-out can be considered as representative for uniform spherical nuclei, even after expansion. 

For a quantitative comparison with the expectations for limiting-temperatures of compound nuclei, 
the region of transition from residue production to multifragmentation 
($Z_{\rm bound}/Z_{\rm proj} \approx 0.7$) seems best suited. 
The residue channels associated with the highest temperatures are found here. 
They may be separated from the fragmentation events in the same bin by applying 
an additional condition on $Z_{\rm max}/Z_{\rm proj}$. 
Furthermore, in order to account for the 
above-mentioned effects of evaporation and to select equivalent degrees of fragmentation, 
slightly lower  
$Z_{\rm bound}$ limits were chosen for the neutron-poor projectiles (by 0.05 on the reduced 
scale, cf. Fig.~\ref{global}). 

\begin{table}
\caption{\label{tab:lnkb}
Limiting temperatures $T_{\rm lim}$ from Ref.~\protect\cite{besp89} for the nominal
isotopes $^{124}$Sn, $^{124}$La, and $^{107}$Sn
and $T_{\rm lim,0.75A}$ for the corresponding nuclei with 75\% of the nominal mass and the 
same $A/Z$ in comparison with the experimental double-isotope 
temperatures $T_{\rm HeLi}$ and $T_{\rm HeLi,res}$, taken as 120\% of the 
apparent values at 
$Z_{\rm bound} /Z_{\rm proj}$ intervals [0.6,0.8] for $^{124}$Sn and
[0.55,0.75] for the neutron-poor cases. For $T_{\rm HeLi,res}$, 
the additional condition $Z_{\rm max}/Z_{\rm proj} \ge 0.6$ 
(0.55 in the neutron-poor cases) was applied. 
All values are given in MeV, the errors are purely statistical.
}
\begin{ruledtabular}
\begin{tabular}{l c c c c}
Projectile & $T_{\rm lim}$ & $T_{\rm lim,0.75A}$ & $T_{\rm HeLi}$ & $T_{\rm HeLi,res}$  \\
\hline
\\
 $^{124}$Sn & ~8.2 & 9.2 & ~6.27 $\pm$ 0.04 & ~5.96 $\pm$ 0.08  \\
 $^{124}$La & ~6.3 & 7.6 & ~5.89 $\pm$ 0.05 & ~5.59 $\pm$ 0.11  \\
 $^{107}$Sn & ~6.6 & 8.2 & ~5.79 $\pm$ 0.05 & ~5.22 $\pm$ 0.09  \\
\end{tabular}
\end{ruledtabular}
\label{table}
\end{table}

In Table~\ref{table}, a summary of the experimental values for these event classes
is given together with the Hartree-Fock results of Besprosvany and 
Levit~\cite{besp89}. Besides the predictions for the nominal projectiles, also
those for nuclei with the same $A/Z$ but only 75\% of the projectile mass are included.
These are the spectator systems most likely populating this bin~\cite{poch95}. 
Their limiting temperatures are higher than those of the nominal nuclei while their difference is
slightly smaller. 

The displayed experimental temperatures contain a 20\% side-feeding correction.
The additional condition on $Z_{\rm max}$ reduces the mean experimental temperature by 
$\Delta T = 0.4 \pm 0.1$~MeV (last column of Table~\ref{table}). Less violent processes 
associated with smaller energy deposits are selected~\cite{poch95}. 
With the same condition, also the difference between the neutron-rich 
and neutron-poor spectator systems changes, rising by a small amount 
from $\Delta T = 0.40 \pm 0.05$ to $0.6 \pm 0.1$~MeV (statistical errors).

It is obvious that Coulomb effects on the scale of MeV as exhibited by the Hartree-Fock 
limiting temperatures are not observed. On the other hand, the large difference in average 
magnitude of the predicted and measured temperatures is not as crucial as it may appear at first 
sight. As noted already by Natowitz et al.~\cite{nato95}, the predictions depend sensitively 
on the type of force used in the calculations~\cite{baldo04}. 
The experimental average $T_{\rm HeLi,res} \approx 5.6$~MeV
for $A \approx 90$ nuclei is close to the results obtained with the SkM$^*$ force by 
Song and Su~\cite{song91} who, however, have not studied the dependence on $A/Z$. 
If these low values, including the corresponding critical temperature $T_c \approx 14$~MeV 
for infinite nuclear matter~\cite{song91}, can be shown to be realistic 
a link may be established between the limits of dynamic compound stability and the onset of 
multifragmentation. Otherwise, and as suggested by the SMM results~\cite{ogul02}, also the 
transition to multifragmentation is predominantly governed by the properties of the fragmentation 
phase space. 

In summary, the study of projectile fragmentation over wide ranges of $A/Z$, up to presently
available limits for proton-rich beams, has shown that the overall isotopic dependence is weak.
In particular, the breakup temperatures entering the nuclear caloric curve were found to be 
identical within a few hundreds of keV, compatible with the assumption of identical 
reaction trajectories usually made in isotopic reaction studies, and in good agreement with the 
SMM predictions for a statistical population of the asymptotic phase space including the 
partition degree of freedom.
The temperature differences reach a value of about 0.6 MeV in selected channels of residue 
production. The temperature $T \approx 5.6$~MeV measured for these processes 
represents a lower bound for the limiting temperature of compound nuclei in the 
$A \approx 90$ region. 

The authors thank J. B. Natowitz for valuable comments and discussion.
C. Sf. acknowledges the receipt of an Alexander-von-Humboldt fellowship.
This work has been supported by the European Community under contract No. HPRI-CT-1999-00001 
and by the Polish Ministry of Science and Higher Education under Contracts No. 1 P03B 105
28 (2005 - 2006) and N202 160 32/4308 (2007-2009).

\end{document}